\documentclass[a4paper,twocolumn,groupeaddress,superscriptaddress,showpacs,cite,aps,pra]{revtex4-1}
 \pdfoutput=1 
\usepackage{epsfig,times,pslatex}
\usepackage{graphicx}
\usepackage{amsmath}
\usepackage{subfigure}
\usepackage{amssymb}
\usepackage[english]{babel}
\usepackage[T1]{fontenc}
\usepackage[latin1]{inputenc}
\usepackage{setspace,latexsym, revsymb,bm}
\usepackage{bbm,bm}

\begin{document}

\newcommand{\Figref}[1]{Fig.~\ref{#1}}

\title{Isospin Correlations in two-partite Hexagonal Optical Lattices}
\author{Marta Prada} 
\thanks{These two authors contributed equally}
\author{Eva-Maria Richter}
\thanks{These two authors contributed equally}
\author{ Daniela Pfannkuche}
\affiliation{I. Institut f\"ur Theoretische Physik, Universit\"at Hamburg, Jungiusstr. 9, 20355 Hamburg, Germany}

\begin{abstract}
Two-component mixtures in optical lattices reveal a rich variety of different phases.
We employ an exact diagonalization method to obtain the relevant correlation functions in  hexagonal optical lattices 
to characterize those phases.  
We relate the occupation difference of the two species to the magnetic polarization. 
`Iso'-magnetic correlations disclose the nature of the system, 
which can be of easy-axis type, bearing phase segregation, or of easy-plane type, corresponding to super-counter-fluidity. 
In the latter case, the correlations reveal easy-plane segregation, involving a highly-entangled state. 
We identify striking correlated supersolid phases appearing within the superfluid limit. 
\end{abstract}
\maketitle
\section{Introduction}
Ultracold quantum gases in optical lattices have long been recognized as a field of growing interest
owing to the possibility to prepare, manipulate and control a quantum system, hence {\em quantum engineering}
\cite{bloch02,zoller98, lewenstein07, zwerger08, bloch08,metcalf}.  
The idealities in optical lattices have indeed been extensively employed as quantum simulators for many-body physics 
\cite{esslinger11,esslinger08,esslinger06,luehmann08,stamper-kurn12,sachdev91,porras11}. 
Multicomponent systems provide, for example, an ideal platform on which to study magnetic ordering \cite{hofstetter09, altman03}. 
In particular, multicomponent graphene-like structures are boasting a profound impact in condensed-matter physics owing not 
only to the fascinating properties of graphene itself, but also to 
the rich quantum phases they exhibit \cite{guinea,soltan,lee,uehlinger, becker10, soltan12}. 

In this  work we explore the extended phase diagram of a two-component hexagonal optical  lattice. 
Most of the theoretical studies employ approximate methods to describe the Mott-insulator regime 
\cite{mathey09,HofstetterNew,Hofstetter12,kuklov03,lukin03,altman03,yang13}, %
or perform expansions around the superfluid regime \cite{powell09}. 
On the contrary, we address in the present work the phase diagram on its full range, from weak to strong coupling, by employing an exact diagonalization method.  
We aim for the iso-magnetic correlations beyond the weak coupling limits, 
and explore in particular the superfluid regime.   
Such technique is of particular significance, since quantum correlations are central to understanding 
spin ordering and non-local string ordering \cite{endres11}.  
We identify ferro- and anti-ferromagnetic correlated regimes, where we find striking supersolid \cite{prokof'ev07,Kuklov11,Hofstetter08,keilmann09} 
and highly-entangled super-counter-fluid (SCF) phases \cite{mathey11pra,mathey09SS,kuklov03}, respectively.

\section{Methods}
Our numerical simulations are based on the single band Bose-Hubbard-Model \cite{zoller98} 
for a two component Bose mixture (boson `A' and boson `B') with nearest neighbor  (NN) hopping, 
which we describe by a Hubbard Hamiltonian, 
\begin{eqnarray}
\label{eq1}
 {H} &= & H_{\rm o} + H_{\rm hop},  \\  
\label{eqos}
 H_{\rm o} & = &\sum_{i} \left\{ \frac{U}{2}\left[\hat{n}_{A}^i(\hat{n}_{A}^i-1)+\hat{n}_{B}^i (\hat{n}_{B}^i-1)\right]+ V_{}\hat{n}_{A}^i\hat{n}_{B}^i\right\},\\
\label{eqhop}
H_{\rm hop}&= & -t \sum_{\langle ij\rangle_1} \left (\hat{a}_{i}^{\dagger} \hat{a}_{j}+\hat{b}_{i}^{\dagger} \hat{b}_{j} + {\rm c.c.}\right).  
 \end{eqnarray}
Here, $i$ runs over the $N$ sites of our unit cell, $t$ describes the hopping for both species,  $U$, $V_{}$ are the on-site repulsion
for equal and different species, respectively, 
$\hat{a}$, $\hat{a}^{\dagger}$ ($\hat{b}$, $\hat{b}^{\dagger}$) 
are the annihilation and creation operators, respectively, for the A (B) particle, and  
 $\hat{n}_{\rm A}=\hat a^{\dagger}\hat a$ ($\hat{n}_{\rm B}=\hat b^{\dagger}\hat b$) the particle number operators.  
 $\langle ij\rangle_1$ indicates that the sum is performed over first NN.
The Hamiltonian of Eq. (3)  suggest expressing the Hilbert space  in terms of  occupation numbers, 
being spanned by all possible configurations of a fixed $N$ and fixed fillings 
$\langle n_\alpha^i\rangle$ = 1/2, $\alpha$ = A, B, 
yielding the holonomic constraint $\langle \hat n^i_{\rm A} + \hat n^i_{\rm B}\rangle = 2S$, with $S = 1/2$. 
We evaluate all observables in the ground state of (\ref{eq1}), $|\psi_{\rm GS}\rangle$, 
which we calculate employing an exact diagonalization method (Lanczos algorithm) with periodic boundary conditions. 
Throughout this work, we employ the notation 
$\langle \mathrm{O} \rangle \equiv \langle \psi_{\rm GS}  |\mathrm{O}|\psi_{\rm GS} \rangle $. 

The on-site term can be expressed, omitting trivial constant terms, as 
\begin{equation}
H_{\rm o} = 
U\left[ (\hat S_x^i)^2 + (\hat S_y^i)^2 + 2(\hat S_z^i)^2\right]
+ 
V \left[ (\hat S_x^i)^2 + (\hat S_y^i)^2\right],
\label{eqSs}
\end{equation}
where we have employed the iso-spin,  $\hat{ S}^{i}_{z}=(\hat n^{i}_{\rm A}-\hat n^{i}_{\rm B})/2$, 
within the Schwinger boson representation \cite{schwinger}: 
$\hat S_\pm^i = \hat S_x^i \pm i \hat S_y^i$, with 
$\hat S_+^i = \hat a^\dagger_i\hat b_i$ and  $\hat S_-^i = \hat b^\dagger_i\hat a_i$. 
In view of Eq. (\ref{eqSs}), 
a first approach to understanding the magnetic phases 
is by evaluating the on-site square of the spin components, 
$\langle \hat S^2_z \rangle \equiv \sum_i\langle (\hat S^i_z)^2 \rangle/N$ and 
$\langle \hat S^2_{\parallel} \rangle \equiv \sum_i[\langle(\hat S^i_x)^2+ (\hat S^i_y)^2\rangle]/N$, 
allowing to classify the ground state as easy-plane or easy-axis type. 

It is customary to distinguish  between the weak coupling, low mobility limit, 
also termed as `Mott insulator' (MI) and a high-mobility, or superfluid regime (SF). 
The SF is thus characterized by a wavefunction broaden over the entire lattice, as opposed to localized 
at the lattice sites, which occurs in the MI regime. 
Both limits can be distinguished by the mean particle fluctuations, 
\begin{equation}
\label{eqD}
\langle \Delta \hat{n}_{} \rangle =\sqrt{\langle (\hat{n}_{\rm A}+\hat{n}_{\rm B})^{2}\rangle-\langle \hat{n}_{\rm A}+\hat{n}_{\rm B}\rangle^{2}} = 
\sqrt{4\langle \hat S^2 \rangle- 3 } . 
\end{equation} 
Naturally, $\Delta \hat{n}_{}$ is expected to be small (large) in the MI (SF) regime. 
We stress that our exact diagonalization method is  not limited to the MI regime in this work, but rather we explore 
the entire parametric region. 
We also note that in the superfluid phase, the usual restriction to spin $S=1/2$ states per site can not be applied in the SF regime. 

It is well known that in the weak coupling, MI limit,  $ t< V_{},U$ 
the hopping term of Eq. (\ref{eqhop}) can be mapped onto an effective iso-spin Hamiltonian 
by a Schrieffer-Wolff-transformation \cite{kuklov03,lukin03}, resulting in the anisotropic Heisenberg spin-1/2 model, 
\begin{equation} 
H_{\rm hop}=  \frac{t^2}{2} \left(\frac{1}{U}-\frac{1}{V_{}}\right)
\left(\hat S_{-}^{i}\hat S_{+}^{j}+\hat S_{+}^{i}\hat S_{-}^{j} - \hat S_{z}^{i}\hat S_{z}^{j}\right).  
\label{eq2}
\end{equation}
However, we are also interested 
in the SF limit, where finite multiple occupation makes the mapping complicated to visualize. 
To explore the magnetic ordering in any regime, we introduce the easy-axis pair-correlation function,  
\[ g^z_{ij} =\langle \hat{n}_{\rm A}^i\hat{n}^j_{\rm A}+\hat{n}^i_{\rm B}\hat n^j_{\rm B} 
-\hat{n}_{\rm A}^i\hat{n}^j_{\rm B}-\hat{n}^i_{\rm B}\hat{n}^j_{\rm A}\rangle.\] 
The first two terms can be  identified with ferromagnetic (FM) contributions and the last two, with 
anti-ferromagnetic (AFM) ones. 
 $g^z_{ij} = {\rm FM} - {\rm AFM}$, 
is thus an observable that quantifies the  magnetic ordering, being (anti)-ferromagnetic when positive (negative). 
In view of Eq. (\ref{eq2}), we expect a FM order in the limit $V_{}> U\gg t$,  
since the prefactor of $\hat S^i_z\hat S^j_z$  is positive. 
Likewise, one may expect AFM ordering in the $U_{}>V\gg t$, where the prefactor changes sign. 
As we will see below, however, in the SF limit $U_{}, V\gtrsim t$  entanglement plays a key role in 
determining the nature of the quantum magnetic phases in the SF limit, which are far from being obvious.

Entangled states suggest the exploration of pair-correlation functions also within the easy-plane, 
\[ g^\parallel_{ij}\equiv 
\langle \hat{\vec{ S}}^i_\parallel \hat{\vec{ S}}^j_\parallel \rangle = 
\langle \hat S^i_+ \hat S^j_- + \hat S^i_-\hat S^j_+ \rangle,\] 
with $g_{ij}^\parallel$ being an observable that quantifies the entanglement between sites `$i$' and `$j$'.  
The different correlation functions reveal indeed  a rich phase diagram, as will be seen in the 
next section.

\section{Results} 
We choose a unit cell  larger than the customary hexagonal lattice's unit cell 
for our numerical calculations, since the hopping term breaks 
translational symmetry, which is only restored on a mean field level. 
We first consider the unit cell depicted in the insets of Fig. \ref{fig_onsite}(a) (broken lines), with 
$N$ = 8, commensurate with a super-solid phase (see below). 
The next unit cell in size commensurate with the super-solid while preserving the $D_{6}$ point group symmetry 
of the lattice contains 24 sites, which is beyond the limit of our present computational capabilities.

To investigate the magnetic character, we 
first calculate the square of the on-site out-of-plane (easy axis) and in-plane (easy plane) components of
spin, $\langle \hat S_z^2\rangle $  and 
$\langle\hat S_\parallel^2\rangle$. 
The out-of-plane component is largest in the $V>U$ region, as shown in Fig. \ref{fig_onsite}(a): 
the dark blue colored region, $\langle \hat S_z^2\rangle >1/4$, reflects multiple occupation of single sites with alike particles.
On the contrary, 
Fig. \ref{fig_onsite}(b) demonstrates an easy-plane ground state in the region $V\lesssim 10t$, 
revealing entanglement of bosons of different species rather than just the alternating arrangement of the right inset of Fig. \ref{fig_onsite}(a). 
Easy-plane state implies super counter-flow (SCF) states \cite{kuklov03,mathey11pra, mathey09}. 
We stress that 
the square of the total spin per site is only a `good' quantum number in the limit $U,V\gg t$,  
where $\langle \hat S^2\rangle = 3/4$  [ dark-blue in Fig. 2(a)]. 
\begin{figure}[!hbt]
\parbox[c]{1.\linewidth}
{
\includegraphics[angle=0, width=1\linewidth]{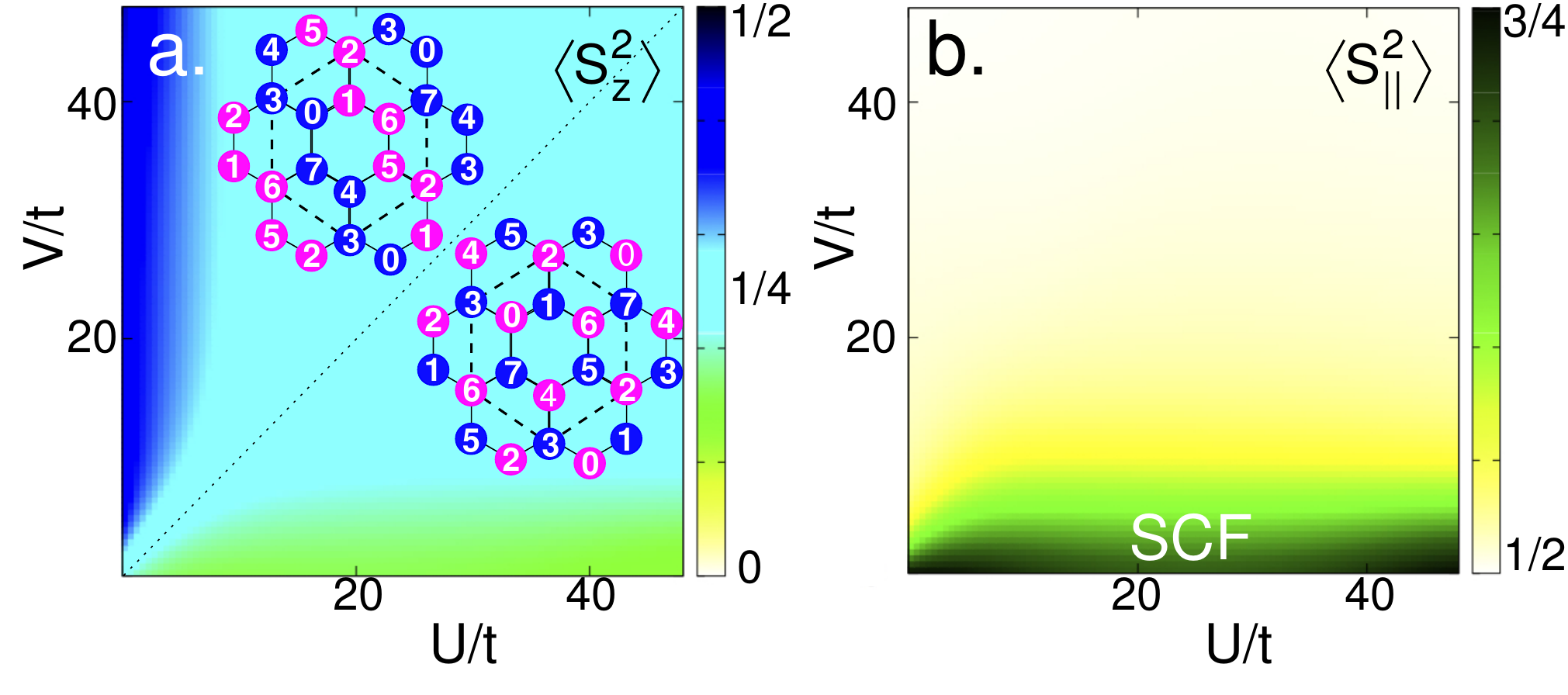}}\\
\caption{
Numerical results for the on-site square of the spin components, easy-axis (a)
 and easy-plane (b). Insets of (a) depicts the FM ($V>U$) and AFM ($U>V$) phase, with the broken lines marking the unit cell boundaries. 
}
\label{fig_onsite}
\end{figure}
The insets of Fig. \ref{fig_onsite}(a) 
illustrate the  MI FM and AFM phases occurring at $V>U$ and $U>V$, respectively. 
Blue and magenta circles encode the two bosonic species, `A' and `B'. 

\begin{figure}[!hbt]
\parbox[c]{1.\linewidth}
{
\includegraphics[angle=0, width=1.\linewidth]{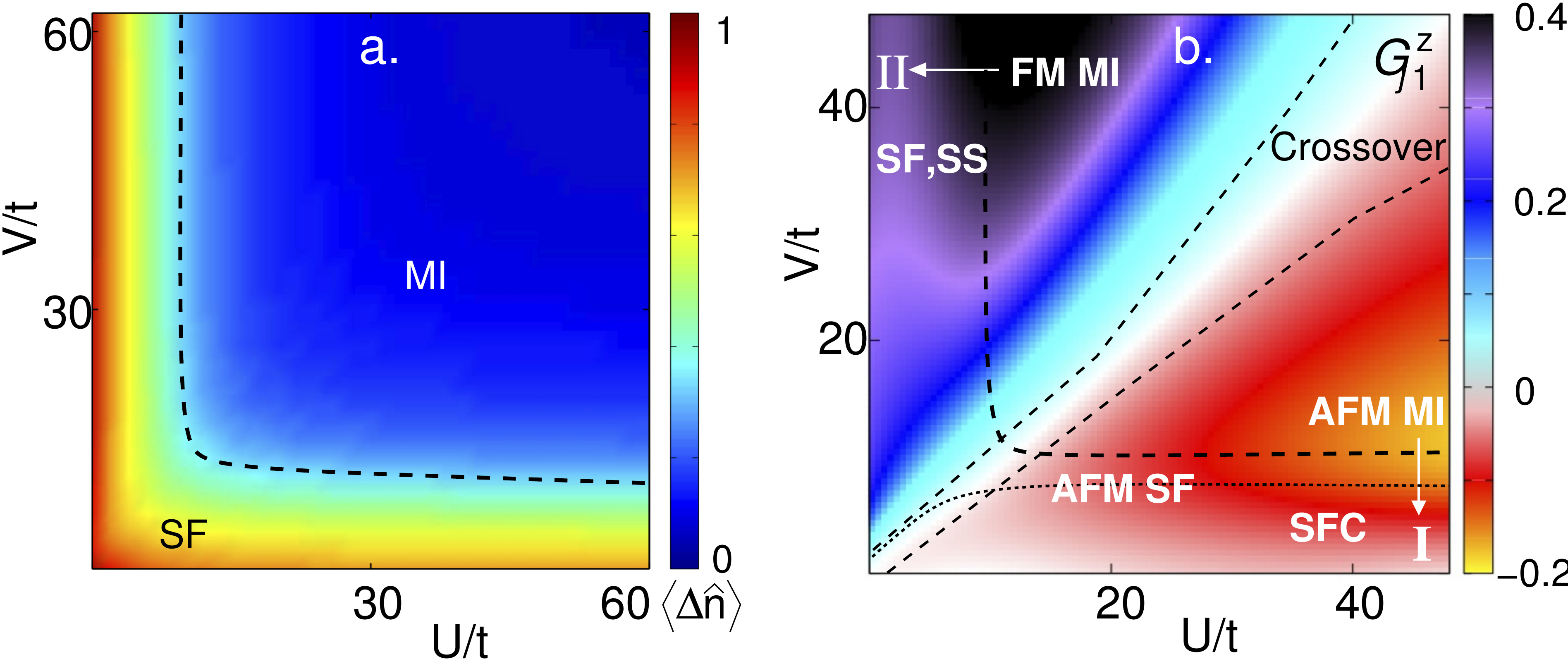}}\\
\caption{
(a) Numerical results for the particle fluctuation $\langle\Delta \hat{n}_{}\rangle$ 
as a function of the on-site interactions, $U$ and $V$. 
A low mobility, MI phase 
(dark blue) and a high-mobility, SF phase (red to cyan) are apparent. 
The dashed line defines the boundary between both limits (arbitrary). 
(b) Numerical results 
 of ${\cal G}^z_1 $ as a function of $U$ and $V$. 
The sign of  ${\cal G}^z_1 $ determines the magnetic phases: FM (cyan-blue-purple-black)  and 
AFM (red-yellow). 
}
\label{fig:phasediagram}
\end{figure}

We now evaluate the averaged particle fluctuations $\Delta \hat{n}_{}$ defined in Eq. (\ref{eqD}).  
The results are presented in Fig. \ref{fig:phasediagram} (a)  as a function of on-site interaction parameters, $V_{}$ and $U$.  
The MI regime (dark blue region) is characterized by low particle fluctuations, with the bosons being localized at 
the lattice sites and hence, well defined $\langle \hat S^2\rangle  = 3/4$. On the contrary, in  the SF regime (rainbow colored region), 
the fluctuations imply $\Delta n >0$ (or $\langle \hat S^2\rangle > 3/4$).
The dashed black curve of Fig. \ref{fig:phasediagram}(a) distinguishes the Mott-insulator 
and the superfluid phases. 
We stress that we do not attempt to determine a sharp, well defined boundary between the different phases throughout this work, 
but rather to determine the overall ordering and magnetic character, by evaluating exact ground states 
of a finite size system. 
Sharp, well defined phase boundaries are expected to occur for certain order parameters in larger systems, for which we are computationally limited.

\begin{figure}[!hbt]
\parbox[c]{1.\linewidth}
{
\includegraphics[angle=0, width=1\linewidth]{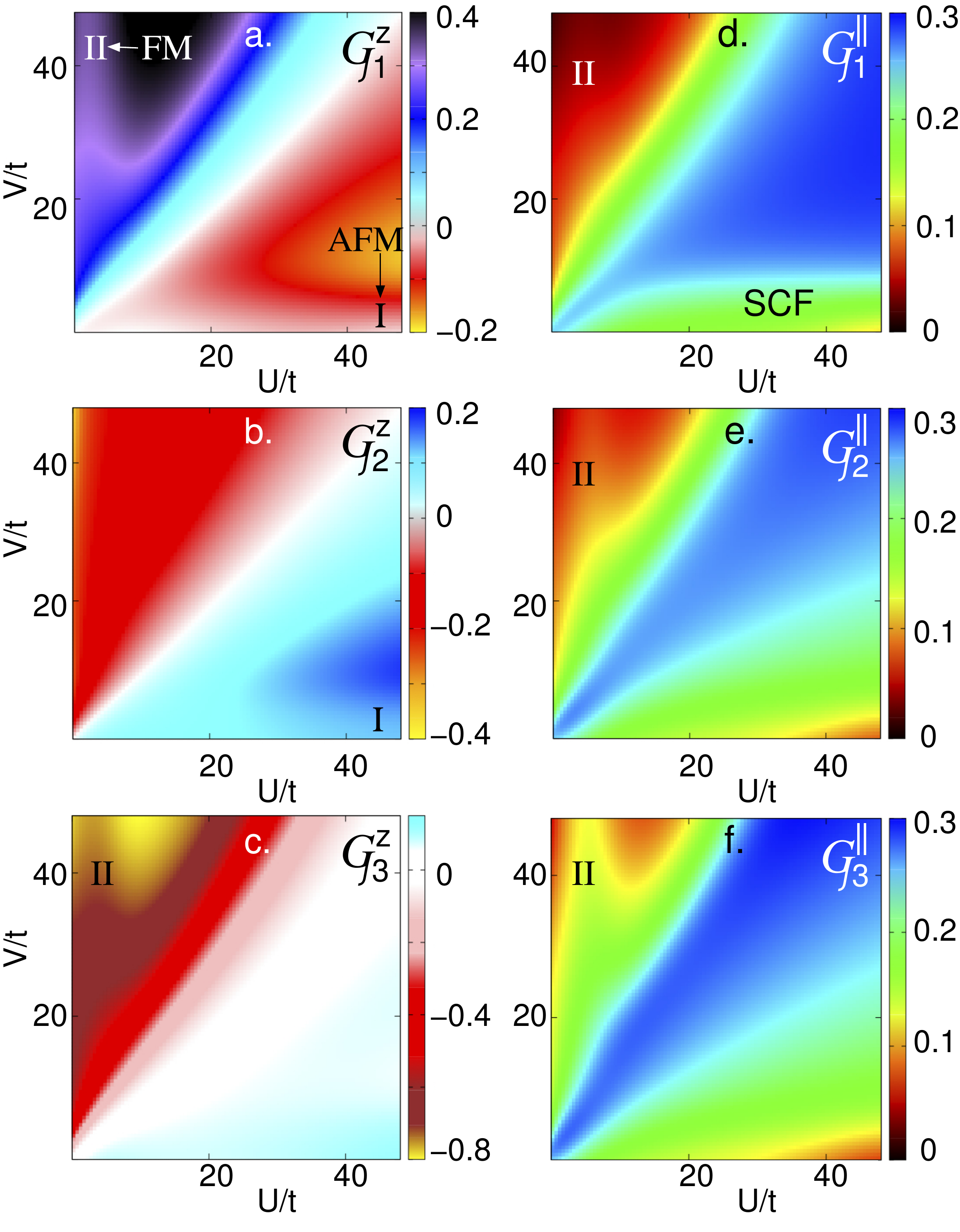}}\\
\caption{ 
Numerical results for easy-axis (a-c) (note the different color scales) and easy-plane (d-e) correlation functions for 
first (a, d), second (b, e) and third (c, f)  NN, for a unit cell with 8 sites.  
}
\label{fig4}
\end{figure}

We next study the magnetic phases by computing the normalized easy-axis pair-density correlation function for NN, 
\begin{equation} 
{\cal G}^z_1 = \sum_{\langle ij \rangle_1} \tilde g^z_{ij} = 
\frac{2n-1}{2 } 
\sum_{\langle i j \rangle_1, \alpha\neq\beta }
\left[ \frac{\langle \hat{n}_{\alpha}^i\hat{n}_{\alpha}^j 
\rangle   }{n-1}  -
\frac{\langle \hat{n}_{\alpha}^i\hat{n}_{\beta}^j 
 \rangle   }{n}   
\right],  
\end{equation}
where $\tilde g^z_{ij}$ is similar to $g^z_{ij}$,  except for correcting factors to take into account 
finite size unit cells \cite{footnote}. 
Fig. \ref{fig:phasediagram}(b) shows the numerical results for ${\cal G}^z_1$ (color scale)
as a function of $U$ and $V_{}$.  
The sign of ${\cal G}^z_1$ discriminates between FM ordering, 
when positive (blue-purple-black), and AFM ordering, when negative (red-yellow). 
In FM ordering ($U$<$V$), `A' and `B' bosons are mainly located in separated domains, whereas  
in 
 AFM ordering ($U$>$V$) the bosons are in an entangled 
state rather than just alternating ordering. 
As a consequence, the absolute value of the magnetic ordering is larger 
in the FM region than in the AFM one [see Fig \ref{fig_onsite}(b)]. 
At the crossover region $U\simeq V$, the ferro- and anti-ferromagnetic correlations 
have the same magnitude, owing to the inherent SU(2) symmetry (white), with  `A' and `B' bosons being close to 
indistinguishable. 
Fig. \ref{fig:phasediagram} (b) reveals a region of large FM correlations (black) and a region of large 
AFM correlations (yellow), both shifted from the vertical ($U=0$) and horizontal ($V=0$) axis, respectively. 
These `shifts' [see arrows in Fig. \ref{fig4}(a)] imply that (I) lowering $U$ at constant $V$ (increasing the $V/U$ ratio), the FM  
character of the system decreases and (II) lowering $V$ at constant $U$ (increasing the $U/V$ ratio), the AFM character decreases, 
both being in principle counterintuitive.  
These anomalies in ${\cal G}_1^z$  suggest exploring second and third NN correlations,  
 ${\cal G}_{2(3)}^z$, 
 as well as the analogous `in-plane' correlations, ${\cal G}_n^\parallel$, 
defined as 
\[
{\cal G}_n^\parallel = \frac{1}{\Omega_n}\sum_{\langle ij\rangle_n}  
 \langle \hat S^i_+ \hat S^j_- + \hat S^i_-\hat S^j_+ \rangle, 
\] 
with $\Omega_n$ being the normalization factors, $\Omega_{n} = \eta_n(n_{\rm A} + n_{\rm B})/2$, 
with $\eta_n$ being the number of $n$-th NN,  and the sum  being performed for  $n$-th NN. 

Fig. \ref{fig4} shows the numerical results for the easy-axis (a-c) and easy-plane (d-e) 
correlations for 1st (a, d), 2nd (b, e) and 3rd (c, f) NN. 
${\cal G}_1^z$  is replotted here for clarity and completeness. 
It is worth noting the different color scales used to plot Fig. \ref{fig4} (a-c), to enhance the relevant features. 
We first focus on the phase at `I' [lower right of Fig. \ref{fig4}(a)].
The feature is reproduced in ${\cal G}_2^z$ [Fig. \ref{fig4} (b)]. 
In the MI limit, alternating ordering enhances the  out of plane correlations, 
which naturally are FM to second NN. 
Recalling the results of Fig. \ref{fig_onsite}(b), we conclude that the spin lies mostly in 
the easy-plane at this region, bearing the  reduction on ${\cal G}_{1,2}^z$ at phase `I' with respect to the deep MI AFM phase. 
Multiple occupation in this phase within the SF regime is limited to two 
particles of different species (a third particle would be penalized, as it would imply two 
particles of same kind), implying doubly-occupied sites alternating with  empty ones. 
This is reflected in  the reduction of the in-plane correlations with respect 
to the AFM phase [lower right of Fig. \ref{fig4} (d)]. We conclude that phase `I' corresponds to a easy-plane 
state with in-plane FM ordering. 

\begin{figure}[!hbt]
\parbox[c]{1.\linewidth}
{
\includegraphics[angle=0, width=1.\linewidth]{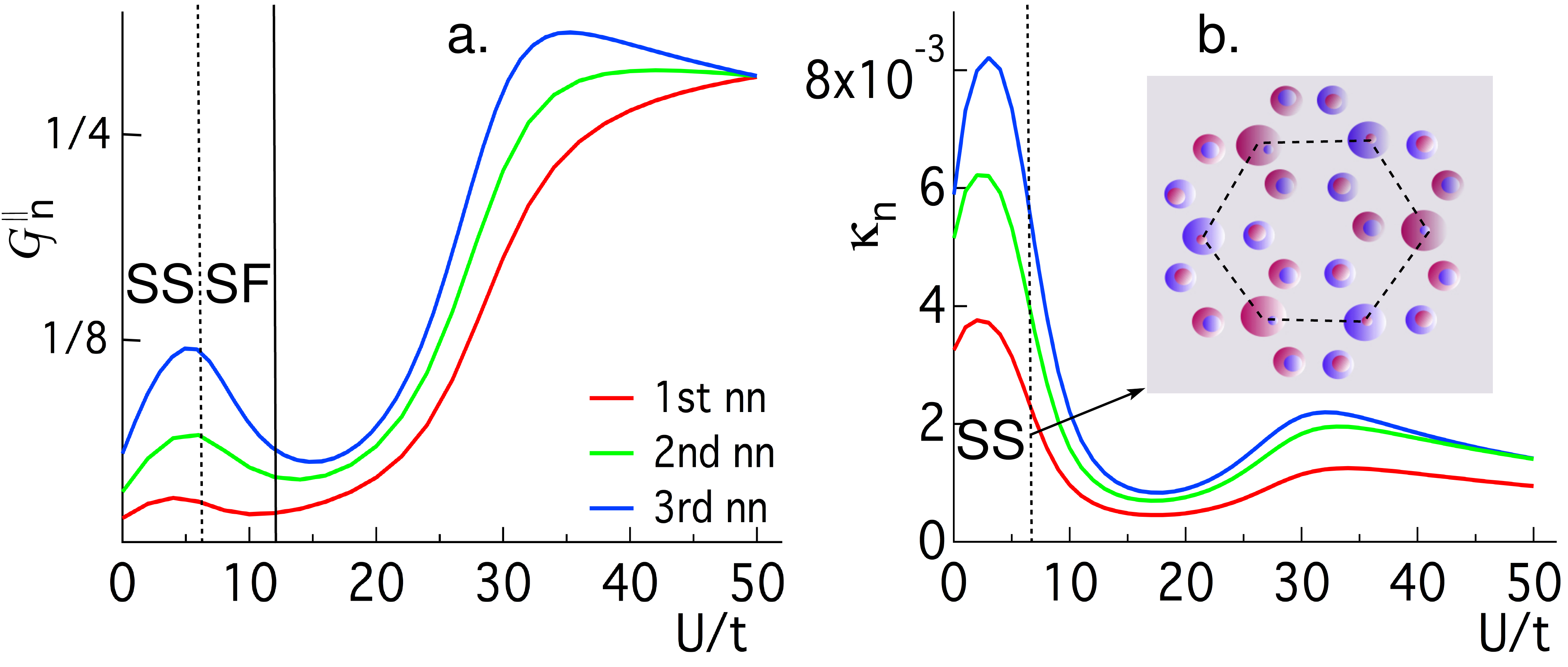}}\\
\caption{
 First (red), second (green) and third (blue) NN  
(a) easy-plane pair-correlation function, ${\cal G}_n^\parallel$ and (b) 
doubly-occupancy flip-flop pair-correlation function, $\kappa_n$. 
All the calculations where performed  with $V = 50t$. Inset: super-solid configuration. 
The black broken lines mark the unit cell boundaries. 
}
\label{figV50}
\end{figure}

We now focus on the phase marked `II' [upper left of Fig. \ref{fig4}(a)]. 
The feature is reproduced in  ${\cal G}_3^z$ [see Fig. \ref{fig4}(c)]. 
It is worth noting that ${\cal G}_3^z$ indicates AFM character where ${\cal G}_1^z$ is FM, 
which is due to the finite size of the unitary cell.
In the $V\gg U$ and within the SF regime, multiple-occupied states of same bosonic species would be expected, 
separated by the  maximum distance $d$ = 3 (recall that the unit cell has eight atoms). 
However,  ${\cal G}_n^\parallel$ reveals a more complex situation [upper left of Fig. \ref{fig4}(d--f)], with
the easy plane correlation functions  peaking around this `anomalous' region.  
For clarity, we present in Fig. \ref{figV50}(a) the $V/t = 50$ resuts for ${\cal G}_{n}^\parallel$, where the 
enhancemnets of the correlation functions are apparent. 
We identify the phase at `II' with the formation of a super-solid, where  a density wave of multiple occupied 
states alternate with entangled states, hence with a larger unit cell (see inset of Fig. \ref{figV50}b).


To demonstrate the formation of a super-solid phase, we further explore higher order correlation functions, 
as: 
\[
\kappa_{n} = \frac{1}{\Omega_n} \sum_{\langle i j \rangle_n}\langle\hat n_{\rm A}^i(\hat n_{\rm A}^i-1)(\hat S_-^i\hat S_+^j + \hat S_+^i\hat S_-^j)\rangle,
\] which can be viewed as a `flip-flop' involving a triple-occupied site with at least two bosons of the same species. 
The correlation function $\kappa_{n}$ for first ($n=1$, red), second ($n=2$, green) and third 
($n=3$, blue) NN is plotted in Fig. \ref{figV50} (b). 
The three functions peak around the super-solid formation.  
The super-cell represented in the inset of Fig. \ref{figV50} (b): the blue and red  balloons 
represent the `A' and `B' character, respectively. 
The black lines are a guide to the eye, and indicate the super-cell formation, 
where $\kappa_{n}$ peaks for all $n$. 

\begin{figure}[!hbt]
\parbox[c]{1.\linewidth}
{
\includegraphics[angle=0, width=1.\linewidth]{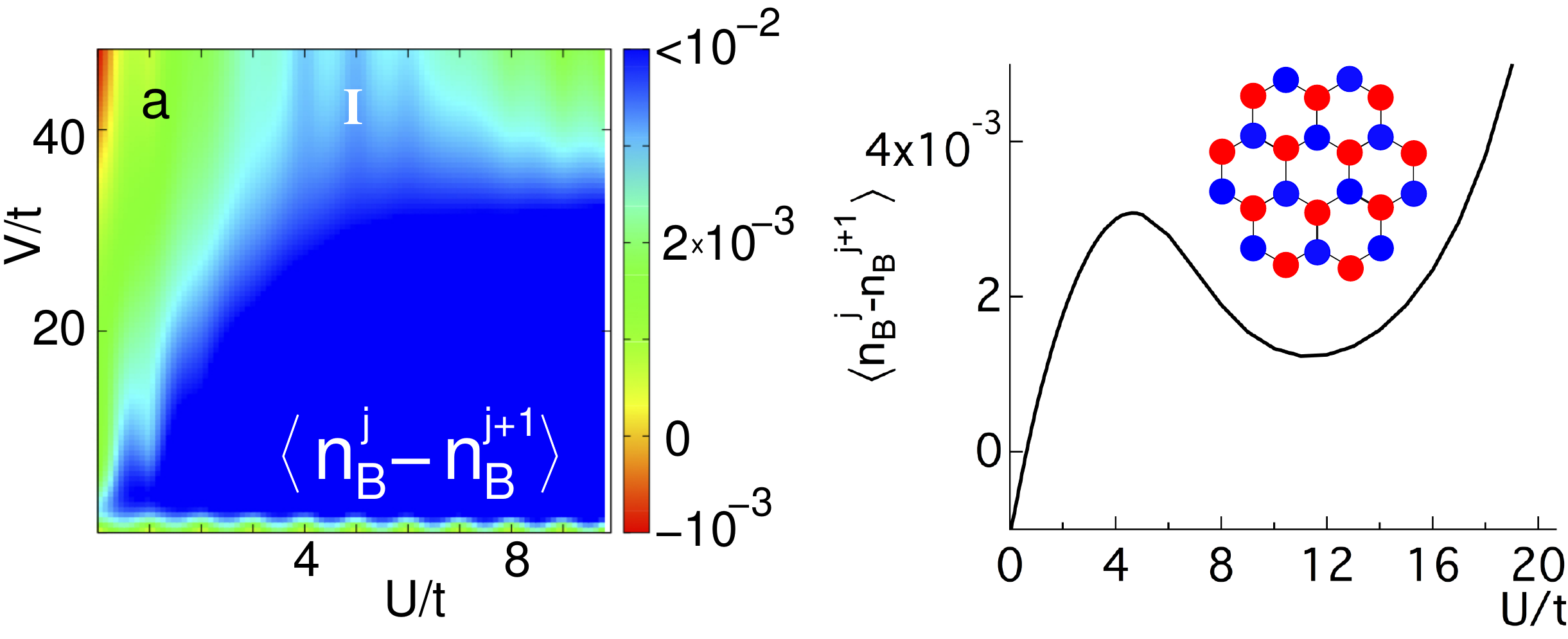}}\\
\caption{ 
Expectation value of NN population difference $\langle n_{\rm B}^i - n_{\rm B}^{i+1}\rangle$
calculated for $\epsilon_A = 0.1t$ as a function of $U$ and $V$ (a) and for $V=50t$, as a function of $U$ (b). 
Inset: $V^\prime$ defines two alternating sublattices (red and blue). 
}
\label{fig_SB}
\end{figure}

We explore further the unconventional magnetic ordering by introducing a spin-dependent lattice 
potential to our Hubbard Hamiltonian, 
\[
\widehat V^\prime = \epsilon_A \sum_{i^\prime} \hat n_{\rm A}^{i^\prime},  
\]
where $i^\prime$ indicates that the sum is performed over alternating sites 
(colored blue in the inset of Fig. \ref{fig_SB}). 
$\widehat V^\prime$ leads to two sublattices, which can be experimentally implemented by employing 
laser beams with defined polarization \cite{becker10, soltan12}.
Introducing the symmetry breaking potential $\widehat V^\prime$, the super-solid will be pinned and 
can be detected using easy-axis correlation functions, hence experimentally accessible. 
To this end, we compute the ground state of $\widehat H + \widehat V^\prime$, for $\epsilon_A=0.1t$,
and consider the NN population imbalance, $\langle n_{\rm B}^i - n_{\rm B}^{i+1}\rangle$. 
The population imbalance is expected to be maximal in the AFM regime, where the `B' atoms would be localized in the `$i^\prime$' 
sublattice,   decreasing towards the FM regime up to the SF regime. In the SF regime, the population imbalance should remain low and roughly constant. 
Fig. \ref{fig_SB} (a) shows, however, a population imbalance peak within the  SF limit (marked I). 
We identify this with the super-solid  signal, where multiply `B'-occupied sites appear surrounded by (dominantly) B-type NN
[see inset of Fig. \ref{figV50}(b)]. 
Note that the $U$ range for this figure differs from the previous ones, as we focus in the super-solid region. 
Fig \ref{fig_SB} (b) is a slice taken at $V = 50t$, where the super-solid signal can be appreciated at nearly the same 
parametric regime as the ones for $\kappa_n$. 
The inset depicts the two sublattices (red and blue).

\begin{figure}[!hbt]
\parbox[c]{.95\linewidth}
{
\includegraphics[angle=0, width=0.8\linewidth]{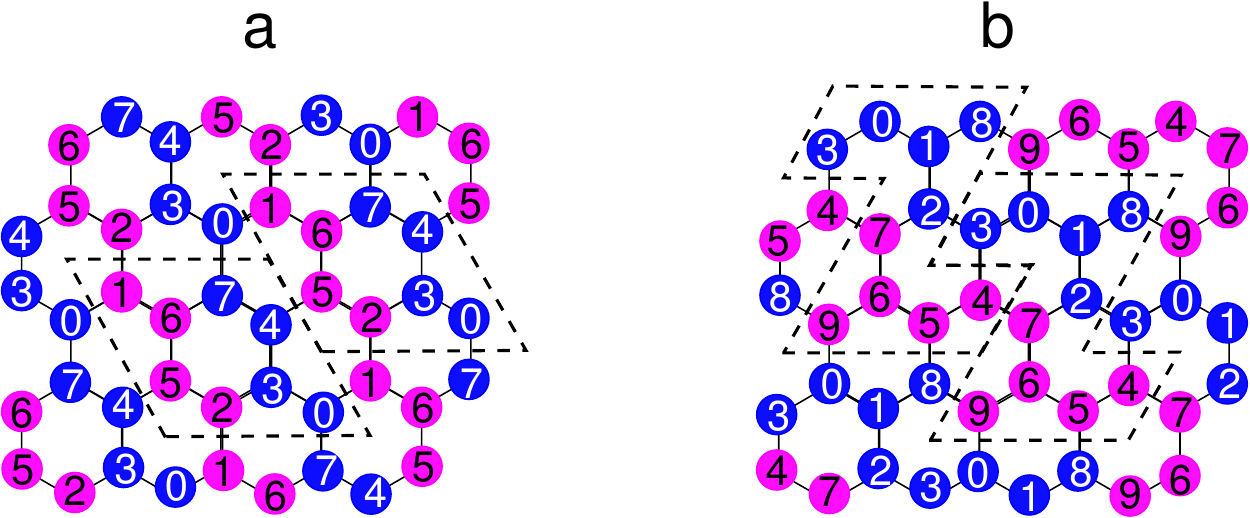}}\\
\caption{
FM phase of (a) 8 and (b) 10 sites unit cell.  The former is a stripe phase, whereas the latter presents phase separation. 
The broken lines depict the boundaries of two unit cells in either case. 
}
\label{fig_lattice}
\end{figure}

In order to account for finite size effects, 
we consider now a unit cell of ten sites within our hexagonal lattice structure. 
Fig. \ref{fig_lattice} (a) depicts the $N$ = 8 FM stripe phase, whereas
Fig. \ref{fig_lattice} (b) exhibits a clear domain structure indicating phase separation within the FM ordering for a $N$ = 10 unit cell. 
We stress that $N  =10$ is not commensurate with the SS structure previously observed. 

\begin{figure}[!hbt]
\parbox[c]{1.\linewidth}
{
\includegraphics[angle=0, width=1.\linewidth]{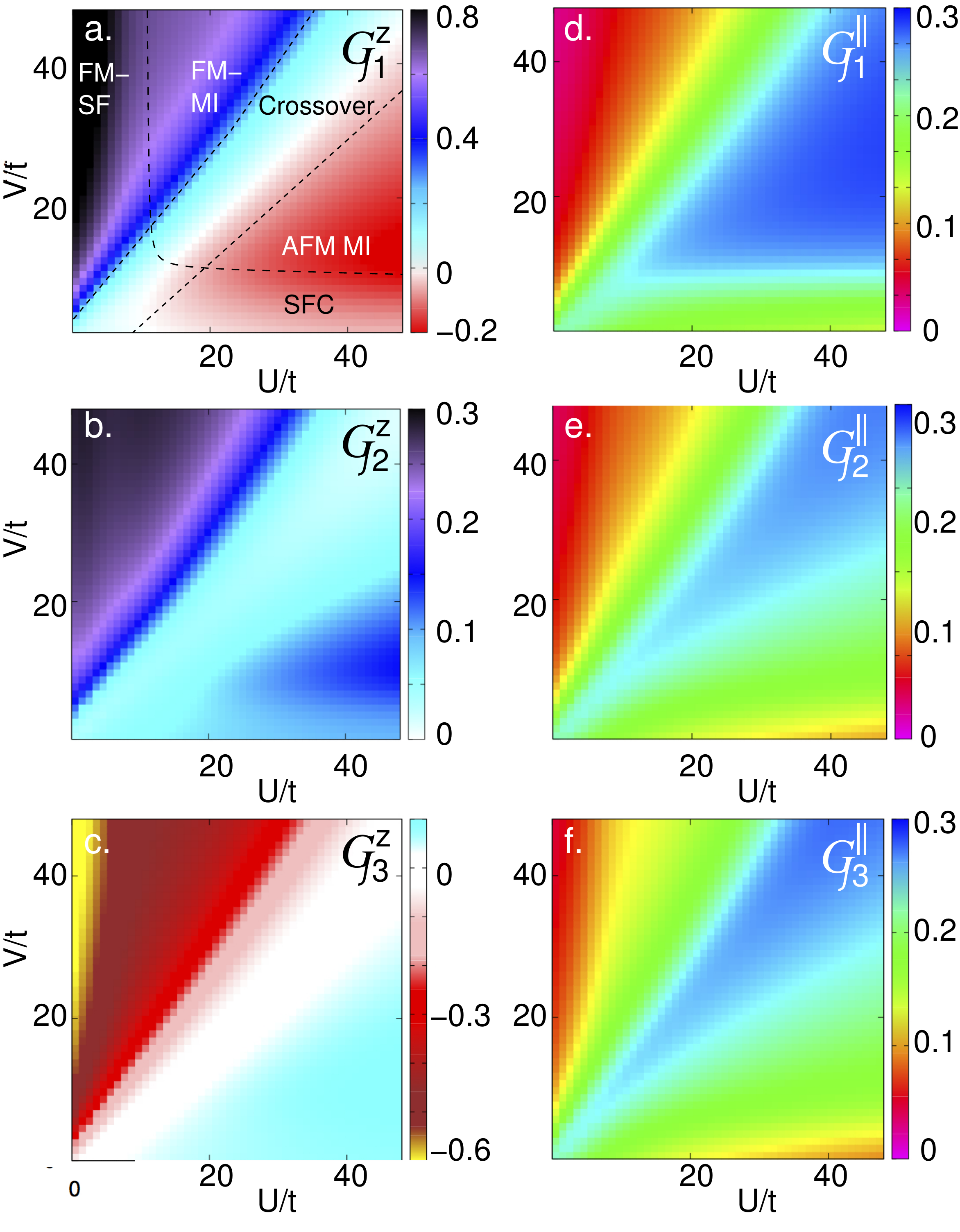}}\\
\caption{ 
Numerical results for easy-axis(a-c) and easy-plane (d-e) correlation functions for
first (a, d), second (b, e) and third (c, f)  NN,  for a 10 sites unit cell. 
Note the different color scales used in (a-c).
}
\label{fig10s}
\end{figure}

The numerical results obtained for the ten sites unit cell 
where consistent with those obtained for the eight site unit cell, reproducing quite well the results presented 
so far.  
Indeed, Fig. \ref{fig_onsite} was reproduced exactly for the $N$ = 10 case (not shown). 
Fig. \ref{fig10s}(a) demonstrates a similar phase diagram as the one in Fig. 3 (b), where the main differences are 
the phase separation and the absence of SS phase, both enhancing the FM ordering (note the different color scale). 
Fig. \ref{fig10s}(b) 
shows FM ordering in the second NN correlation, consistent with the phase separation depicted in Fig. \ref{fig_lattice} (b). 
We point out a small FM third NN correlations within the AFM regime in Fig. \ref{fig10s}(c) (cyan), which appeared also in Fig. \ref{fig4}(c).  
In that region, quantum fluctuations would wash out the magnetic character, 
however, our normalization \cite{footnote} favours the FM phase, explaining the small FM correlations.
Fig. \ref{fig10s}(c-f) shows similarities with  Fig. 4(c-f), respectively, except for the absence of the super-solid signal. 
We would like to stress again that the absence of super-solid phase is due to the incommensurability of the super-solid periodicity 
with the periodic boundary conditions for the 10 sites unit cell. 
It is worth noting that the asymmetry of  ${\cal}G_{1}^\parallel$ [Fig. \ref{fig10s}(d) and \ref{fig4}(d) ] 
with respect to the $U=V$ line demonstrates a `canted' spin phase within the AFM regime.
Note also the equivalent scales in Fig. \ref{fig10s}(d-f) and Fig. 4(d-f), showing also large in-plane 
correlations in the AFM regime. 
These results suggest a long range FM ordering within the easy-plane component of the spin,  
in contrast to the short range out of plane AFM ordering. 

 



\section{Conclusions}
In summary, we calculated the extended zero-temperature quantum phase diagram of a two component 
Bose gas in a hexagonal lattice for MI, SF and intermediate limit. 
Employing an exact diagonalization method, we obtained a classification scheme based on pair correlation functions.  
A rich phase diagram is observed, in particular in the SF regime. 
Two-operator correlation functions allowed to distinguish not only between FM and AFM phases,  
but also between  easy-axis and easy-plane ground states, bearing phase separation and SCF, respectively. 
Non-trivial fourth-operator correlations demonstrate the formation of a super-solid phase.  
Signatures of the super-solid phase could be identified in the 
population imbalance by artificially breaking the symmetry with a spin-dependent lattice potential, 
suggesting a feasible experimental method.
We discriminate finite size effects by recalculating the correlations for a unit cell incommensurate 
with the super-solid. 
Easy-axis correlations for $N$ = 8 and $N$ = 10 sites show qualitative and quantitative resemblance, 
and hence, can be regarded as representative for larger systems too. 
Despite the small system size, the correlation functions allow us to discriminate the different phases 
occurring in macroscopic systems.

\section{acknowledgments}
We would like to thank L. Mathey, M. Potthoff, A. Chudnovskiy, and H. Niehus for valuable discussions. 
This work has been supported by the DFG SFB 925. 

\bibliographystyle{plain}
\bibliography{main}

\begin{thebibliography}{5}
\bibitem{footnote}{Noting that if a boson type $A$ occupies site `$i$', 
there are only $n_{\rm A}-1$ remaining bosons to contribute to the FM correlations  
whereas $n_{\rm B}$ particles may contribute to the AFM ones,  
we define the summ of our pair-correlation function as
${\cal G} = \alpha [{\rm FM}/(n-1)  - [\rm AF]/n] $, with $2n = n_{\rm A} + n_{\rm B}$ and 
$\alpha = (n -1/2)/( {\rm FM }+{\rm AF})$. 
}.
\end{thebibliography}

\end{document}